\documentclass[twocolumn,10pt,a4paper,showpacs,aps]{revtex4}
\usepackage{amsfonts,amssymb}
\usepackage{dcolumn}
\usepackage{amsmath}
\usepackage{latexsym}
\newfont{\cyr}{wncyr10 at 9pt}
\newfont{\cyrit}{wncyi10 at 9pt}
\newfont{\cyrbf}{wncyb10 at 9pt}
\begin{document}
\title{Field strength for graded Yang-Mills theory}%
\author{K. Ilyenko${}^*$}%
\affiliation{Institute for Radiophysics and Electronics, NAS of
             Ukraine, 12~Ak.~Proskura Street, Kharkiv - 61085, Ukraine}%
\date{23 July 2003}%
\begin{abstract}
\noindent The field strength is defined for %
\textit{osp}(2/1; $\mathbb{C}$) non-degenerate graded Lie algebra. %
We show that a pair of Grassman-odd scalar fields find their place
as a constituent part of the graded gauge potential on the equal
footing with an ordinary (Grassman-even) one-form taking values in
the proper Lie subalgebra, \textit{su}(2), of the graded Lie
algebra. Some possibilities of constructing a meaningful
variational principle are discussed.
\end{abstract}
\maketitle
\section{INTRODUCTION}
\label{Intro}%
As well-known the number of gauge bosons of a theory is given by
the number of generators of the corresponding gauge Lie algebra.
The requirement of being a Lie algebra stems from the reality of
the action functional which follows from the properties of the
Fermat principle in optics and the Feynman path integral. In
particular, this leads to physically admissible evolution of a
system with such an action functional. Nevertheless, no one is
restricted from looking for not necessarily of Lie-type algebras
as the gauge algebras provided they are physically meaningful.
\section{GRADING IN GAUGE ALGEBRA}
\label{AlgGr}%
Such an algebra is advocated in the current contribution. This is
a graded extension of \textit{su}(2) gauge algebra by a pair of
odd generators, $\tau_A$, which anticommute with one another and
commute with the three even generators,  $T_a$, of \textit{su}(2).
We use the square brackets to denote both commutation and
anticommutation operations of the generators with understanding of
their proper usage. The defining relations have the form,
\cite{Kac1977,Hughes1981,Brooks1996}:
\begin{eqnarray}
&[T_{a}, T_{b}] = i\varepsilon_{abc}T_{c},\;\;\; [T_{a}, \tau_{A}]
= \frac{1}{2}(\sigma_a)_{A}^{\hphantom{A}B}\tau_{B},& \nonumber \\
&[\tau_{A}, \tau_{B}] = \frac{i}{2}(\sigma^a)_{AB}T_{a}.&
\label{DefRel}
\end{eqnarray}
Lowercase Roman indices run from 1 to 3; uppercase Roman indices
run over 1, 2; $(\sigma_a)_{AB}$ =
$(\sigma_a)_A^{\hphantom{A}C}\epsilon_{CB}$; $\varepsilon_{abc}$
($\varepsilon_{123}$ = 1) and $\epsilon_{AB}$ ($\epsilon_{12}$ =
1) are the Levi-Civita totally antisymmetric symbols in three and
two dimensions; the Pauli matrices $(\sigma_a)_A^{\hphantom{A}B}$
and $\epsilon_{AB}$ are given by
\begin{displaymath}
\mbox{{\tiny %
$(\sigma_a)_{A}^{\hphantom{A}B} = (\sigma^a)_{A}^{\hphantom{A}B} =
\left[ \left(
\begin{array}{cc}
0 & 1 \\ 1 & 0
\end{array}
\right),
\left(
\begin{array}{rl}
 0 & i \\
-i & 0
\end{array} \right),%
\left( \begin{array}{cc} 1 &\! 0 \\ 0 &\! -1
\end{array}
\right) \right], %
\epsilon_{AB} = \left(
\begin{array}{cc}
\!\! 0 & 1 \\
    -1 & 0
\end{array}
\right).%
$}}%
\end{displaymath}
The non-degenerate super Killing form,
$B(T_\alpha, T_\beta)$ , is defined by
\begin{equation}
B(T_\alpha, T_\beta) = \frac{2}{3}\,\mbox{str}(T_\alpha T_\beta) =
\left(\mbox{%
\begin{tabular}{c|c}
$\delta_{ab}$& $0$ \\ \hline
          $0$& $i\epsilon_{AB}$ \\
\end{tabular}}
 \right),
\label{sKillingF}
\end{equation}
where the supertrace operation is adopted from \cite{Cornwell1989}
and the Greek indices from the beginning of the alphabet run over
the whole set of the graded Lie algebra generators. It turns out
that all of the generators are grade star Hermitian: on the even
ones being just Hermitian in an ordinary sense while the odd
generators obey more complicated relations (cf.
\cite{SNR19771,SNR19772}). We assign a degree,
$\mbox{deg}T_\alpha$, 0 to the even and 1 to the odd generators.
\section{GAUGE POTENTIAL}%
Given an \textit{su}($N$) Lie algebra one defines a gauge
potential, which takes values in the algebra, by introducing $(N^2
- 1)\times\,n$ one-forms $A_\mu^a(x)\mbox{d}x^\mu$, $n$ being the
dimension of space-time, and transvecting them with the algebra
generators  $T_a$. Note that from the present standpoint we have a
composite object of the degree (0,1), the first position shows
that generator $T_a$ is, by definition, an even element of the Lie
algebra and the second position exhibits that
$A_\mu^a(x)\mbox{d}x^\mu$ is a one-form in the algebra of exterior
differential forms on space-time. This has a suggestive
generalization to the case when one also has the degree 1 odd part
of a graded Lie algebra: (s)he needs to construct the homogeneous
compliment of the expression above, namely, the element of degree
(1,0). It has the form $\tau_A\mathit{\Phi}^A(x)$ and must be
added to the element of degree (0,1) to form the complete graded
gauge potential
\begin{eqnarray}\label{gConn}
\mathcal{A}(x) & = & T_aA_\mu^a(x)\mbox{d}x^\mu +
\tau_A\mathit{\Phi}^A(x)
\\%
homogeneous & = & even\otimes{}odd + odd\otimes{}even \nonumber
\end{eqnarray}
Here $\mathit{\Phi}^A(x)$ are zero-forms on space-time. Thus one
obtains a proper element $\mathcal{A}(x) =
T_\alpha{}A^\alpha{}(x)$ [$A^\alpha{}(x) \equiv
(A^a_{\mu}\mbox{d}x^\mu, \mathit{\Phi}^A)$] of the total degree
one in the direct product of two graded algebras (cf.,
\cite[Eq.~(2.45)]{CNS1975} and \cite[p.~629]{Lang1985}).
\section{GRADED FIELD STRENGTH}%
The graded field strength, $\mathcal{F}(x)$, is defined by means
of taking the exterior derivative of the graded gauge potential,
$\mathcal{A}(x)$, and adding its wedge product with itself,
$\mathcal{A}(x)\wedge\mathcal{A}(x)$ multiplied by the interaction
constant:
\begin{equation}\label{gFStrength}
\mathcal{F}(x) = \mbox{d}\mathcal{A}(x) + (\mbox{i}g/2)
\mathcal{A}(x)\wedge\mathcal{A}(x).
\end{equation}
Now we shall clarify the meaning of both operations. The first
term on the right-hand side of (\ref{gFStrength}) is given by
\begin{eqnarray}
\mbox{d}\mathcal{A}(x) = \frac{1}{2!}T_a[\partial_\mu{}A^a_\nu(x)
\hspace{-0.7ex} & - & \hspace{-0.5ex}
\partial_\nu{}A^a_\mu(x)]\mbox{d}x^\mu\wedge\mbox{d}x^\nu \nonumber \\
\hspace{-0.5ex} & + &
\hspace{-0.5ex}\tau_A\mbox{d}\mathit{\Phi}^A(x). \label{Lterm}
\end{eqnarray}
The second term on the right-hand side of (\ref{gFStrength}) needs
more explanations. First, as in the case with Yang-Mills theory,
one defines the wedge product for algebra valued forms
$\mathcal{A}(x)$. Second, we have to deal with the odd part of
graded Lie algebra and zero-forms involved in the graded gauge
potential. Thus we define, (cf., \cite[Eq.~(2.45)]{CNS1975} and
\cite[p.~632]{Lang1985}),
\begin{eqnarray}\label{NLterm}
 \mathcal{A}\wedge\mathcal{A} & = & (T_\alpha\otimes{}A^\alpha)\wedge
 (T_\beta\otimes{}A^\beta) \\
 & \equiv & (-1)^{\mbox{deg}A^{\alpha}\mbox{deg}T_\beta}[T_\alpha,
 T_\beta]\otimes{}A^\alpha\wedge{}A^\beta, \nonumber
\end{eqnarray}
where $[T_\alpha, T_\beta]$ is a commutator if at least one of the
graded Lie algebra generators $T_\alpha$ is even and an
anticommutator otherwise and $\mbox{deg}A^\alpha$ means the degree
in the exterior algebra of differential forms on space-time. One
can easily see that this straight generalization is bilinear, goes
into the usual Yang-Mills result for ordinary Lie algebras and
makes possible the graded Jacobi identity for the direct product
of two graded algebras: the exterior and matrix ones.

Finally, we unite (\ref{Lterm}) and (\ref{NLterm}) in one
expression:
\begin{equation}\label{FSbad}
\mathcal{F} = \frac{1}{2!}T_a[F^a_{\mu\nu}\mbox{d}x^\mu\wedge
\mbox{d}x^\nu + (\sigma^a)_{AB}\mathit{\Phi}^A\mathit{\Phi}^B] +
\tau_A\varphi^A,
\end{equation}
where we has denoted
\begin{equation}\label{OFStrength}
\varphi^A(x) = \mbox{d}\mathit{\Phi}^A(x) +
\frac{1}{2!}(\sigma_a)_B^{\hphantom{B}A}\mathit{\Phi}^B(x)A^a_{\mu}(x)
\mbox{d}x^\mu
\end{equation}
and $F^a_{\mu\nu}(x) = \partial_\mu{}A^a_\nu(x) -
\partial_\nu{}A^a_\mu(x) - g\varepsilon_{abc}A^b_\mu(x)A^c_\nu(x)$
is just the usual Yang-Mills field strength. We intend to preserve
the grading of the gauge potential in the expression for the field
strength. The second summand in (\ref{FSbad}) is of the degree
(1,1), the expression
$T_aF^a_{\mu\nu}\mbox{d}x^\mu\wedge\mbox{d}x^\nu$ is of the degree
(0,2), while the expression
$T_a(\sigma^a)_{AB}\mathit{\Phi}^A\mathit{\Phi}^B$ being of the
degree (0,0) falls out of this pattern. Allowing for the symmetric
property of the Pauli matrices $(\sigma^a)_{AB}$ =
$(\sigma^a)_{BA}$, we shall assume that the functions
$\mathit{\Phi}^A(x)$ are the Grassman-odd quantities, which obey
the anticommutation relation $\mathit{\Phi}^A\mathit{\Phi}^B$ =
${}-\mathit{\Phi}^B\mathit{\Phi}^A$. Then, equation (\ref{FSbad})
takes the form
\begin{equation}\label{FSgood}
\mathcal{F} = \frac{1}{2!}T_aF^a_{\mu\nu}\mbox{d}x^\mu\wedge
\mbox{d}x^\nu + \tau_A\varphi^A.
\end{equation}
Thus, we have built a homogeneous expression for the graded field
strength of the total degree two, as expected.
\vspace{1em}
\section{DISCUSSION AND OUTLOOK}%
The very possibility of defining the graded potential and field
strength opens up a number of further questions. Firstly, a
problem of graded gauge invariance arises. We hope that existing
in the literature (see, e.g. \cite{OgMiz1975}) introduction of
Grassman-odd transformation parameters for odd generators of the
graded Lie algebra could provide appropriate solution to this
problem. Secondly, as mentioned at the beginning of the current
contribution, one is interested in a definition of a real-valued
Lagrangian density. This would lead to physically acceptable
Euler-Lagrange equations. In particular, we are going to consider
an invariant with respect to the graded Lie algebra automorphisms
expression
\begin{equation}\label{AP}
    S = \int\mbox{str}(\mathcal{F}\wedge{}^*\mathcal{F})
\end{equation}
as the relevant action functional and explore the corresponding
equations of motion.
\vspace{1em}
\section*{ACKNOWLEDGEMENTS}
\noindent I would like to thank Yu.P.~Stepanovsky and
V.~Pidstrigach for many helpful  discussions.
\begin{center}
{\cyrbf Napryazhennostp1 polya dlya
graduirovanno}$\check{\mbox{\cyrbf
i}}$ \\ {\cyrbf teorii J1nga-Millsa}%
\end{center}

{\cyr Opre\-de\-lya\-et\-}{\cyr sya gra\-dui\-ro\-van\-naya
na\-prya\-zhen\-nostp1 po\-lya dlya} \textit{osp}(2/1;
$\mathbb{C}$) {\cyr ne\-vy\-rozh\-den\-no\symbol{'032}
ka\-li\-bro\-voch\-no\symbol{'032} al\-geb\-ry. Po\-ka\-za\-no,
chto para nechetnykh grassmanovykh po\-le\symbol{'032} mo\-zhet
yavlyatp1sya sostavno\symbol{'032} chastp1yu
gra\-du\-iro\-van\-no\-go 4-po\-ten\-tsia\-la kalibrovochnogo
polya naravne s obychno\symbol{'032} (grassmanovo
chetno\symbol{'032}) odin-formo\symbol{'032}
prinimayuwe\symbol{'032} znacheniya v maksimalp1no\symbol{'032}
sobstvenno\symbol{'032}} \textit{su}(2)-{\cyr podalgebre Li
graduirovanno\symbol{'032} algebry} \textit{osp}(2/1;
$\mathbb{C}$). {\cyr Obsuzhdayut}{\cyr sya nekotorye vozmozhnosti
postroeniya na e1to\symbol{'032} osnove fizicheski priemlemogo
variatsionnogo printsipa.}

${}$

\noindent ${}^*$\textit{Electronic address}: kost@ire.kharkov.ua

\end{document}